\documentclass[epj]{svjour}
\usepackage{graphics}
\usepackage{epsfig}

\journalname{The European Physical Journal B}

\newcommand{\be}{\begin{equation}}
\newcommand{\ee}{\end{equation}}
\newcommand{\bea}{\begin{eqnarray}}
\newcommand{\eea}{\end{eqnarray}}

\newcommand{\dr}{\partial}
\newcommand{\nab}{\vec \nabla}
\newcommand{\ccc}{$\mbox{\textsf{C}}_{\mbox{\textsf{c}}}^{\mbox{\textsf{c}}}$}

\begin{document}

\title{On the crescentic shape of barchan dunes}
\author{Pascal Hersen}
\institute{
Laboratoire de Physique Statistique de l'ENS,
24 rue Lhomond, 75005 Paris, France.\\}
\date{\today}
\abstract{Aeolian sand dunes originate from wind flow and sand bed
interactions.  According to wind properties and sand availability,
they can adopt different shapes, ranging from huge motion-less star
dunes to small and mobile barchan dunes.  The latter are crescentic
and emerge under a unidirectional wind, with a low sand supply.  Here,
a $3d$ model for barchan based on existing $2d$ model is proposed. 
After describing the intrinsic issues of $3d$ modeling, we show that
the deflection of reptating particules due to the shape of the dune
leads to a lateral sand flux deflection, which takes the mathematical
form of a non-linear diffusive process.  This simple and physically
meaningful coupling method is used to understand the shape of barchan
dunes.}
\PACS{ {45.70.-n}{Granular systems} \and {47.54.+r}{Pattern selection;
pattern formation} }
\authorrunning{P. Hersen}
\titlerunning{On the crescentic shape of barchan dune}
\maketitle
\section{Characteristics of Barchan Dunes}
R.A. Bagnold opened the way to the physics of dunes with his famous
book: \emph{The physics of blown sand and desert dunes}
\cite{B41}.  From then on, a great deal of investigations - laboratory
experiments \cite{B41,ACD01,HDA02}, field measurements
\cite{CWG93,PT90,F59,LS64,N66,H67,LL69,H87,S90,HH98,SRPH01} and
numerical computations
\cite{WG86,NO92,W95,NYA97,S97,BAD99,S01,K02,ACD02} - have been
conducted by geologists and physicists.  In particular, a large amount
of work has been dedicated to the barchan, a dune shaped by the
erosion of a unidirectional wind on a firm ground.\\
\begin{figure}[htbp]
\includegraphics{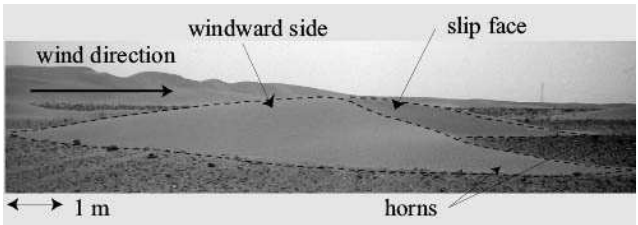} \vspace{-0.1cm} \caption{Side view of a
barchan dune .  The main properties of barchan dunes are outlined: two
horns pointing downwind, the slip-face and the flat main body.  The
barchan shown is approximately $20$ meters long and wide, and $2$
meters high.  The slip-face angle is roughly $30^{o}$ to the vertical,
which corresponds to the angle of repose of a sand-pile.}
\label{fig1}
\end{figure}

\begin{figure}[t]
\includegraphics{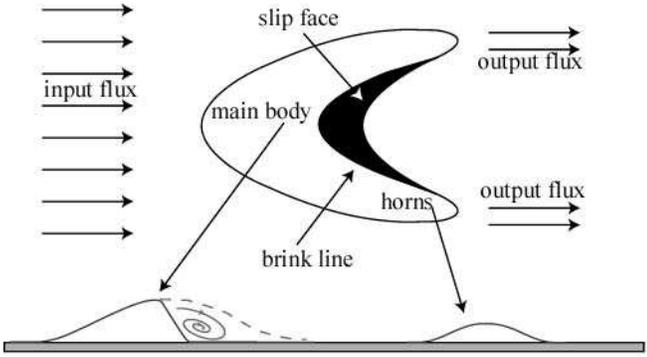} \vspace{-0.1cm} \caption{Barchan dune
properties.  Grains follow the wind direction, and sand flux is not
strongly deflected by the dune relief.  As observed in the field, sand
grains can escape from the horns, but not from the main dune body. 
Instead, they are trapped into the slip-face.  This difference of
behavior between the main body and the horns is the key to understand
the barchans.}
\label{fig2}
\end{figure}

A side view of a barchan - see Fig.~\ref{fig1} - shows a rather flat
aerodynamic structure.  When viewed from above, a barchan presents a
crescentic shape with two horns pointing downwind.  A sharp edge - the
brink line - divides the dune in two areas: the windward side and the
slip face, where avalanches develop - see Fig.~\ref{fig2}.  Because of
a boundary layer separation along this sharp edge
\cite{B41,CWG93,PT90}, a large eddy develops downwind and wind speed
decreases dramatically.  Therefore, the incoming blown sand is dropped
close to the brink line.  That is why the barchan is known to be a
very good sand trapper.  Sometimes, when the drift of sand is too
large, an avalanche occurs and grains are moved down the slip face. 
In short, grains are dragged by the wind from the windward side of the
dune to the bottom of the slip-face and, grain after grain, the dune
moves.  Field observations show that barchans can move up to $70
m/year$ \cite{SRPH01}.  Their speed is dependent on wind power and on
their size: for the same wind strength, the velocity of barchans is
roughly inversely proportional to their heights
\cite{B41,ACD01,CWG93,LL69,S90}.  Barchan dimensions range from $1$ to
$30$ meters high, and from $10$ to $300$ meters long and wide
\cite{B41,CWG93,PT90}.  However, large barchans are often unstable,
leading to complex structures called {\it mega-barchans}
\cite{CWG93,PT90}.  More accurate analyses reveal that the height,
width and length are related by linear relationships \cite{ACD01,HH98}
and that no mature barchan dune smaller than one meter high can be
found: there is a minimal barchan size.  Finally, the last important
characteristic of barchans, but much less well-documented, is the sand
leak at the tip of the horns \cite{B41,CWG93,PT90,HACHD03}, where no
recirculation bubble develops.  This shows that barchan dunes are
three dimensional structures, whose center part and horns have totally
different trapping efficiency - see Fig.~\ref{fig2}.\\

Because of the inherent difficulties of field-work, (a typical mission
duration is rather short compared to the lifetime of a dune and its
shape movement), numerical modeling is, alongside laboratory
experiment \cite{HDA02}, an important method to explore barchan
properties such as their morphogenesis, their stability and their
interaction modes, which are still poorly understood.  Obviously,
these problems are related to the original structure of barchans, and,
accordingly, they should be studied with a $3d$ approach.  The aim of
the present paper is to discuss how to extend existing $2d$ models to
the $3d$ situation.  Then, we will show that the crescentic shape can
be explained by the existence of a sand-flux constraint and a lateral
sand-flux.  In the following part, we start by recalling briefly the
main features of $2d$ modeling of dunes .  Then a model for the $3d$
situation will be presented.
\section{The \ccc\ class of models}
Numerical modeling of dune requires a description of the effect of
wind flow over a sand-bed.  Obviously, computing exact turbulent
numerical solution starting with the {\it Navier-Stokes} equations is
possible, but this would take a very long time.  An alternative is to
use the \ccc\ model \cite{ACD02}, based on the following approach.
\subsection{Numerical model for the $2d$ case}
Let us call $h(x,t)$ the sand bed profile and $q(x,t)$ the vertically
integrated volumic sand-flux.  They are linked by the mass
conservation:
\be
\dr_{t}h + \dr_{x}q = 0 .
\ee
The sand flux $q(x,t)$ is the main physical parameter needed to
understand dune physics.  It cannot locally exceed a saturated value
$q_{sat}(x,t)$, which is the maximum number of grains that the wind is
able to drag per unit of time at $x$.  Previous work have already
focused on the important role played by the saturation of the sand
flux and the so-called saturation length, $l_{sat}$
\cite{HDA02,K02,ACD02}.  The nature of this saturation process can be
understood in terms of sand grain inertia: if the wind speed
increases, it takes some time for the grains, initially at rest, to
reach the wind velocity.  Different approaches are possible to
describe the evolution of the sand flux towards its saturated value
$q_{sat}$.  However for the sake of simplicity, this saturation
process is here taken into account with the equation,
\be
\dr_x q = \frac{q_{sat} - q }{l_{sat}}.
\ee
\noindent This equation keeps the most important effect: the existence
of a characteristic length-scale $l_{sat}$.  Moreover, $q_{sat}$,
depends on the wind shear velocity $u_{*}$.  Even if the nature of
this dependency is still debated, the saturated flux is a growing
function of $u_{*}$ \cite{ASW91,Sorensen01,A03}.  Then a linear
expansion of $q_{sat}$, as proposed in the innovative work of
Sauermann {\it et al.} \cite{S01,K02} from the model of Jackson \& 
Hunt \cite{JH75}leads to:
\be
\frac{q_{sat}(x)}{Q} = 1 + A \int \!\! \frac{d\chi}{\chi} \, \dr_x
h_e(x-\chi) + B \, \dr_x h_e(x),
\ee
\noindent where $Q$ is the saturated flux value on a flat ground and
$h_{e}(x)$ is the envelope of the dune.  This envelope encages the
dune and its recirculation bubble \cite{S90}, and is used to include
the boundary layer separation heuristically.  Notice that in the
latter equation, $h_{e}(x,t)$ appears only through its first spatial
derivative.  It is consistent with the assumption that atmospheric
turbulent boundary layer is fully developed, so changes in $u_{*}$,
and accordingly in $q_{sat}$ are scale-invariant.  From an
aerodynamical point of view, this relation takes into account two
effects: a pressure effect, controlled by $A$, where the whole shape
acts on the wind flow; and a destabilizing effect, controled by $B$,
which ensures that the maximum speed of the flow is reached before the
dune summit.  Although in principle we could compute the parameters
$A$ and $B$, we prefer to consider them as tunable parameters. 
Finally, avalanches are simply taken into account: if the local slope
exceeds a critical value, the sand flux is increased strongly along
the steepest slope.  We will come back to the description of
avalanches in section $4$.  The only scaling quantities are $Q$ and
$l_{sat}$ and they are used to adimensionalize the problem.

In fact, this model belongs to what we called the \ccc\ class.  It
does not depend strongly on the model use for the shear stress: other
models \cite{G02} of shear stress perturbations could be used
providing that they include the role of the whole shape ($A$) and the
asymmetry of the flow ($B$).  The same remarks apply to the charge
equation\cite{S01,K02} and to the equation linking $q_{sat}$ and
$u_{*}$.  This shows the robustness of the physics ingredients used in
this class of model.  Even though this method does not provide the
most detailed results, this kind of model has been used successfully
to model $2d$ barchan profile \cite{S01,K02,ACD02}.
\subsection{Speed dispersion and trapping efficiency}
As a matter of fact, simulations in $2d$ show the existence of two
kinds of solutions: \emph{dune} and \emph{dome} \cite{ACD02}.  The
dune solution has a slip-face that catches all the incoming sand: the
dune can only grow - except if the input sand flux is null.  On the
contrary, for a dome, a large amount of sand can escape, and if the
loss of sand is not balanced by an influx, the dome can only shrink. 
Hence, these two solutions cannot co-exist.  To compute the steady
shape of a $3d$ barchan, it is possible to cut it in "slices" parallel
to the wind direction and, for each slice, to use the $2d$ model to
compute its evolution.  However, in a real barchan, a $2d$ slice from
the horns seems to behave like $2d$ dome solution, while a $2d$ slice
from the main body coresponds to $2d$ dune solution.  However, in
order to make slices from the main body and slices from the horns
coexist, we must introduce a sand flux which will redistribuate
laterally the sand from the center towards the horns.  This sand flux
coupling is also needed to overcome the speed dispersion effect.  As a
matter of fact, if all the slices have initially the same shape - but
at a different scale ratio - the saturated flux at the crest is the
same for all the slices, as imposed by turbulence scale invariance and
the speed of a slice is given by:
\be
  c = \frac{q_{c} - q_{out}}{h_{c}}
\ee
where $q_{c}$ and $q_{out}$ are respectively the flux at the crest and
the output flux, and $h_{c}$ is the height at the crest of the slice. 
Hence, the smaller the slice, the faster its motion.  This dispersion
explains why the barchan takes a crescentic shape.  But to reach a
steady state, all the slices must move at the same speed.  Therefore,
the lateral coupling must also induce a speed homogenization.

\section{Different lateral coupling mechanisms}

What physical mechanisms can lead to a redistribution of the sand flux
on the dune surface?  We can think of three different possibilities:
avalanches, lateral wind shear stress perturbations and grain motions
(saltation and/or reptation).

\subsection{Avalanches}

First, avalanches develop in three dimensions along the steepest
slope.  This creates lateral sand flux in the slip face area. 
However, for real barchan dunes, that sand-transport is directed from
the edges towards the center.  Accordingly, the center part grows and
slows down while the border slices shrink and accelerates.  This does
not constitute a stabilizing mechanism.

\subsection{Wind deflection}

The lateral wind deflection is another possibility.  Field
observations \cite{B41} and numerical simulations \cite{S01} tend to
show that, due to its flatness, the dune does not make the wind flow
deviate too much laterally.  Consequently trajectories of grains are
not dragged into the lateral direction.  Nevertheless, it is always
possible to compute the wind speed perturbations in the lateral
direction and see what it gives.  This has been done recently
\cite{SH03-2}, but there is hardly any evidence that it is the main
physical process responsible for a lateral sand flux on the dune.  On
the other hand, it appears to be an important effect for the
development of lateral instabilities along transversal dunes
\cite{SH03-1}.

\subsection{Saltation and reptation coupling}
\begin{figure}[htbp]
\includegraphics{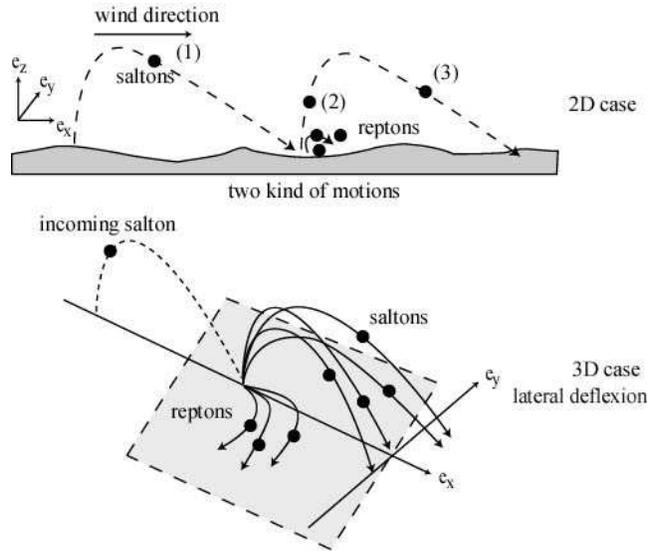} \vspace{0.1cm} \caption{Influence of
gravity.  While trajectories of {\it saltons} are deflected randomly
at each collision, the {\it reptons} are always pushed down along the
steepest slope.}
\label{fig3}
\end{figure}
A better candidate is the grain motion.  Sand transportation can be
described in terms of two species
\cite{A03}: grains in
saltation - the {\it saltons} - and grains in reptation - the {\it
reptons} - see Fig.~\ref{fig3}.  {\it Saltons} are dragged by wind,
collide with the dune surface, rebound, are accelerated again by the
airflow, and so forth.  At each collision, {\it saltons} dislodge many
{\it reptons}, which travel on a short distance, rolling down the
steepest slope, and then, wait for another {\it salton} impact.

As the wind deflection is weak, we assume that {\it saltons} follow
quasi $2d$ trajectory in a vertical plane (see last section), except
when they collide with the dune.  At each collision, they can rebound
in many directions, depending on the local surface properties, and
this induces a lateral sand flux.  Given that the deflection by
collision is strongly dependent on the surface roughness, we assume
that, on average, the deflection of {\it saltons} is smaller than for
{\it reptons}, which are always driven towards the steepest slope.  In
the following derivation, we will neglect the sand flux deflection due
to saltation collisions.

Despite the major role played by {\it saltons} in dune dynamics, the
presence of {\it reptons} should not be dismissed.  According to field
observations \cite{HW88,H77}, {\it reptons} are strongly dependent on
the local slope: this can be observed by looking at the relative
orientation of the wind and ripples.  On hard ground, ripples are
perpendicular to wind direction, but on a dune their relative
orientations change with the local slope \cite{H77}.  Even if one is
reluctant to use the saltation coupling because of its inherent
difficulties, the {\it reptons} would still be rolling on the dune
surface: gravity naturally tends to move the grains along the steepest
slope, leading to lateral coupling of the dune slices.  In the
following part, we will focus on this coupling by reptation, which has
never been used for the study of barchan dunes.  The saltation process
remains important in this model, since it induces reptation coupling.


\section{A  $3d$ model with  reptation coupling}

\subsection{Formulation of the model}

\begin{figure}[htbp]
\includegraphics{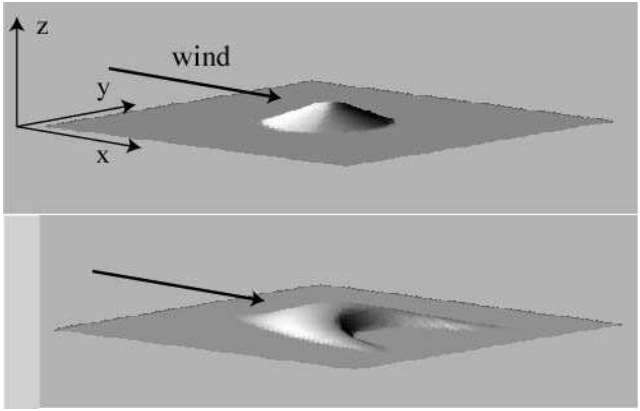} \vspace{-0.1cm} \caption{Typical initial
and final shape of barchan dunes given by the $3d$ \ccc\ model. 
Parameters used are $A = 9.0$, $B = 5.0$, $D = 0.5$ with initially:
$W_{0} = 30 l_{sat}$, $H_{0} = 3 l_{sat}$.}
\label{fig4}
\end{figure}

As {\it reptons} are created by {\it saltons} impacts, the part of
the flux due to {\it reptons} is assumed to be proportional to the
saltation flux, $q_{sal}$ \cite{B41}.  Hence, the flux
of {\it reptons} is simply written as \cite{discussion}:
\be 
\vec{q_{rep}} = \alpha q_{sal}(\vec{e_{x}} - \beta \vec{\nabla} 
h)
\ee 
The $\alpha$ coefficient represents the fraction of the total sand
flux due to reptation on a flat bed.  In case the bed is not flat, the
flux is corrected to the first order of the $h$ derivatives, by a
coefficient $\beta$ and directed along the steepest slope to take into
account the deflection of {\it reptons} trajectories by gravity. 
Assuming that saltation trajectories are $2d$, the total sand flux
$\vec{q}$ is given by:
\be
  \vec{q} = q_{sal} (1+\alpha)\vec{e_{x}}- \beta \alpha q_{sal} 
  \vec{\nabla} h
\ee
Moreover, reptation flux is assumed to instantaneously follow the
saltation flux, so that there is no other charge equation than the
saltation flux one:
\be
\dr_x q_{sal}  = \frac{q_{sat} - q_{sal}}{l_{sat}}.
\ee
\noindent Then, the mass conservation equation becomes:
\be
\dr_t h + \dr_x q_{sal}  +\vec{\nabla}\vec{q_{rep}} = 0 .
\ee
Calling $D = \alpha \beta / (1+\alpha)$ and $\tilde q = (1+\alpha)
q_{sal}$, the two latter equations can be rewritten as :
\be
\dr_x \tilde q = \frac{\tilde q_{sat} - \tilde q}{l_{sat}}
\ee
\be
\dr_t h + \dr_x \tilde q = D ( \dr_x (\tilde q \dr_x h )
+\dr_y (\tilde q\dr_y h) )
\label{eq1}
\ee
Finally, we obtain the same set of equations as in the $2d$ \ccc\
model, but with one more phenomenological parameter, $D$, which can be
understood as the importance of the lateral coupling, because of
lateral deflection of {\it reptons}.  This formulation appears to be a
nonlinear difusion equation driven by the non dimensionnal coefficient
$D$.  For a homogeneous flux solution, $Q D$ appears to be a diffusion
coefficient, showing the diffusion-like role of the coupling
coefficient $D$.  Notice that $\tilde q$ is no longer the saltation
flux, but the part of the flux that does not depend on the bed slope.
Avalanches are computed in three dimensions using a simple
trick.  If the local slope exceeds the threshold $\mu_{d}$, the sand
flux is strongly increased by adding an extra avalanche flux:
\be
\vec{q_{a}} = E (\delta \mu) \nab h,
\ee
where $\delta \mu$ is null when the slope is lower than $\mu_{d}$ and
equal to ($\delta \mu=|\nab h|^2-\mu_d^2$) otherwise.  For a
sufficiently large coefficient $E$, the slope is relaxed independently
of $E$.  Finally, note that quasi-periodic boundary conditions (the
total output flux is reinjected homogeneously in the numerical box)
are used to perform the numerical simulations.  These boundary
conditions are used, first to work with a constant mass, and second to
force the system to converge towards its steady state.  Obviously
these boundary conditions constrain the final shape.

\subsection{The origins of the crescentic shape}
\begin{figure}[htbp]
\includegraphics{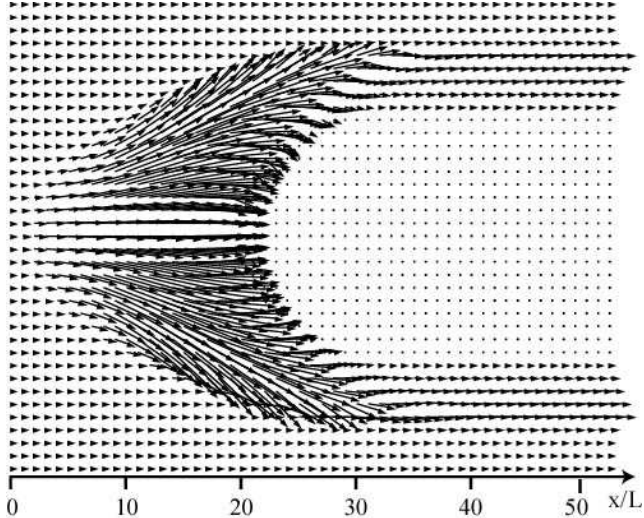} \caption{Lateral diffusion of sand flux. 
The angle of the flux vectors are magnified $3$ times to be clearly
visible.  Both the deflection towards the horns and the presence of
avalanches can be observed.  Parameters used are the same than for
figure \ref{fig4} and \ref{fig6}: $A = 9.0$,$B = 5.0$,$D = 0.5$ with
initially: $W_{0} = 30 l_{sat}$, $H_{0} = 3 l_{sat}$.}
\label{fig5}
\end{figure}
\begin{figure*}[htbp]
\includegraphics{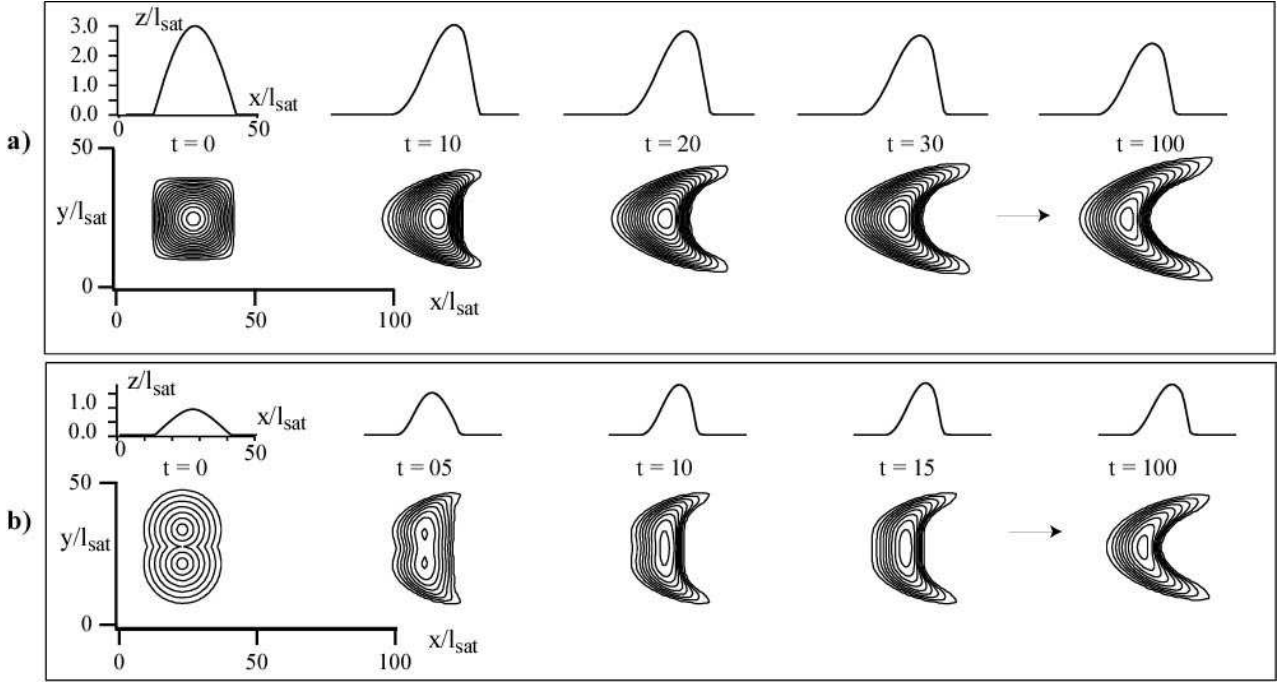} \caption{Barchan formation.  a)
evolution of an initial cosine bump sand pile : $h(x,y) = \cos(2\pi
x/W_{0})\cos(2 \pi y/W_{0})$.  In the beginning, the horns moves
faster than the central part of the dune, leading to the formation of
the crescentic shape.  The shape reaches an equilibrium thanks to the
lateral sand flux, which feeds the horns.  Parameters used are the
same than for figures \ref{fig4} and \ref{fig5}: $A = 9.0$,$B =
5.0$,$D = 0.5$ with initially: $W_{0} = 30 l_{sat}$, $H_{0} = 3
l_{sat}$.  b) Evolution of a bi-cone sand-pile (same parameters than
in (a)).  This shows that emergence of a crescentic shape is
independant on initial shape: the hole in the middle is filled up, and
a crescentic shape forms.  In both boxes the height between two
level-lines is $0.2 l_{sat}$.}
\label{fig6}
\end{figure*}
Our $3d$ \ccc\ model depends on three phenomenological parameters :
$A$ and $B$ take into account aerodynamics effects, and $D$ describes
the efficiency of the slices coupling.  The final shape, if one
exists, depends on these parameters.  For example, Fig.~\ref{fig4}
shows a typical $3d$ final steady state of a computed barchan dune,
which looks likes a real aeolian one.  The arrows on Fig.~\ref{fig5}
indicate the direction of the total sand flux on the whole dune shape. 
The deviation towards the horns is clearly visible as the sand
captured by the slip-face.  Looking at Fig~\ref{fig6} helps us to
understand the formation of the crescentic shape of barchan dune. 
Starting from a sand-pile, some horns will appear and since they are
faster than the center slice of the dune, a crescentic shape appears. 
Now, let us consider a slice and let us call $q_{x}$, the excess flux
due to {\it reptons} coming from the lateral sand flux, $q_{in}$ the
incoming saltation sand flux brought by the wind and $q_{e}$ the flux
due to erosion.  The existence of a saturated flux imposes :
\be
q_{e} + q_{x} + q_{in} < q_{sat}^{max} .
\ee
\noindent where $q_{sat}^{max}$ is the maximum value of the saturated
flux on the given slice.  Thus, if the excess flux $q_{x}$ increases,
the erosion flux, $q_{e}$, decreases.  As the speed of the slice is
governed by the erosion, the slice slows down.  After a while, the
horns receive enough sand from the main body to decrease their speed
and to compensate the output flux.  Finally the speed of all slices is
the same and the barchan moves without changing its shape.  Moreover,
the center part, which would grow without lateral coupling, can now
have an equilibrium shape, since all the extra flux is deviated
towards the horns.  Furthermore this coupling process is stabilizing
with respect to local deformation.  If a slice increases in height,
the excess sand flux leaving the slice will increase, and the
deformation will shrink.  Similarly, starting from a two maxima shape,
the part of the flux sensitive to the local slope tends to fill up the
gap between the two maxima: the whole mass is redistributed and a
single barchan shape is finally obtained - see Fig.~\ref{fig6}.  This
agrees with the apparent robustness of the crescentic shape observed
on the field.  Whatever the external conditions are, the same
morphology is roughly found everywhere where barchans develop.  Hence,
this lateral coupling helps us to understand such barchan structures. 
However, we have no clue about the possible value of this coupling
coefficient $D$ and it is therefore useful to study the influence of
$D$ on the barchanic shape.


\section{Significance of the coefficients}
\subsection{The influence of the coupling coefficient $D$}
\begin{figure}[!h]
\includegraphics{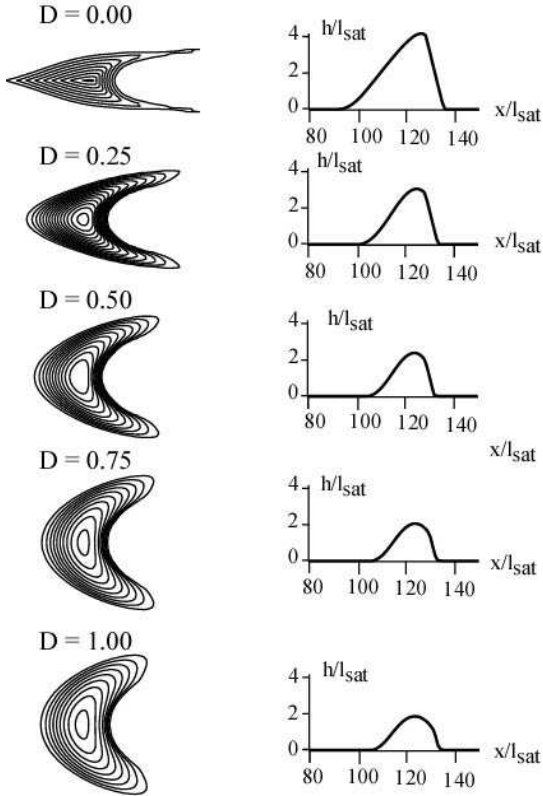} \caption{Different $3d$ shapes obtained
for different values of $D$ and for the same initial sand-pile. 
Parameters used (except for $D$ coefficient) are the same than for
figures \ref{fig4}, \ref{fig5} and \ref{fig6}.  Computations are
performed with quasi-periodic boundary conditions.  The barchan
central slice is represented on the right.  Notice that for $D=0$, a
steady state is reached because of lateral sand redistribution due to
avalanches on the windward side.  However, there is a sharp edge which
separates the dune in two part, leading to an arrow shape rather than
to a crescentic shape.}
\label{fig7}
\end{figure}
As can be seen on Fig.~\ref{fig7}, $D$ has a crucial influence on the
global morphology.  First of all, for a small $D$, ({\it i.e.} a small
lateral sand flux), the flux escaping from the tip of the horns is
mainly due to $q_{in}$ and to the erosion flux, $q_{e}$.  Therefore,
erosion may be important, and the horns may move quickly.  Thus,
during the transient, the barchan elongates.  On the contrary, for a
larger $D$, the lateral flux received by the horns, $q_{x}$, is larger
and then the erosion is less important.  The horns move slower and the
elongation is shorter.\\

\begin{figure}[htbp]
\includegraphics{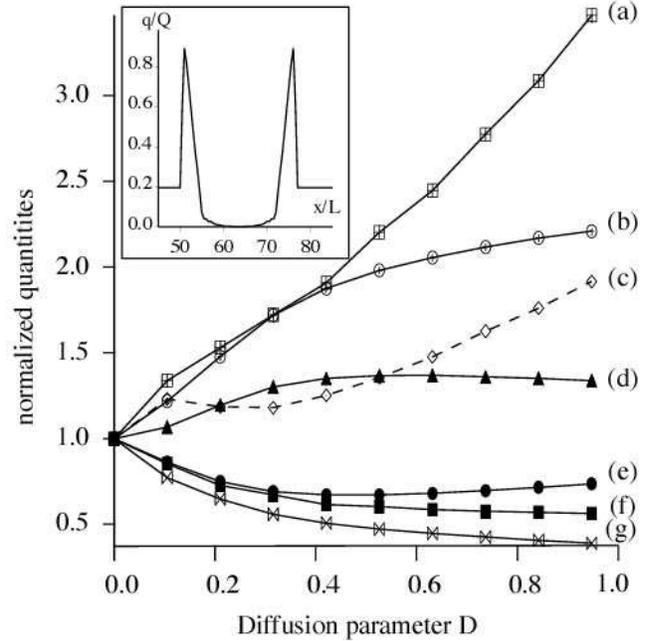} \caption{Dependance of barchan dune
properties on the coupling coefficient $D$.  The case with $D=0$ is
taken has a reference.  The properties are: (a) width of a horn, (b)
total width, (c) equilibrium input flux, (d) speed of the dune , (e)
total length (including horns), (f) length of the center slice, (g)
maximum height.  A typical output flux cross section is shown in the
upper left part of the graph: the width of the horns is measured by
measuring the width of the flux peaks.  The initial sand pile is
always a cosine bump defined by $h(x,y) = H_{0}\cos(2\pi
x/W_{0})\cos(2 \pi y/W_{0})$ with $W_{0} = 20$ and $H_{0} = 2.0$.}
\label{fig8}
\end{figure}

Secondly, the width of the horns is also dependent on $D$.  For a
large $D$, the deflected flux is large.  However, as the output sand
flux cannot exceed a maximum value, horns have to be wide in order to
transport away all the incoming sand, and indeed, the width of the
horns increases with $D$.  This simple explanation is confirmed by
numerical simulation, as depicted in Fig.~\ref{fig7} and
Fig.~\ref{fig8}.  Measuring the width of the horns in the desert may
be a convenient way to estimate the value of $D$ from field
observations.
Finally the aspect ratio of the dune, $h/w$, varies with $D$, because
the coupling mechanism tends to diminish slopes.  In other words, the
dune tends to spread out laterally when $D$ increases, sometimes
leading to structures without a slip-face: $3d$ dome solution.  In
this case, sand can escape from the whole transversal section of the
dune and the output flux is then higher than for the dune solution. 
Basically the transition from $3d$ dune to $3d$ dome situations occurs
when the horn width is equal to half the dune width.  This corresponds
approximately to $D = 2.0$ with the parameters $A = 9.0$ and $B =
5.0$.  For too large a diffusion, the dune becomes very wide and flat,
without a slip-face, and remains unsteady.
\subsection{The influence of coefficients $A$ and $B$}
\begin{figure}[htbp]
\includegraphics{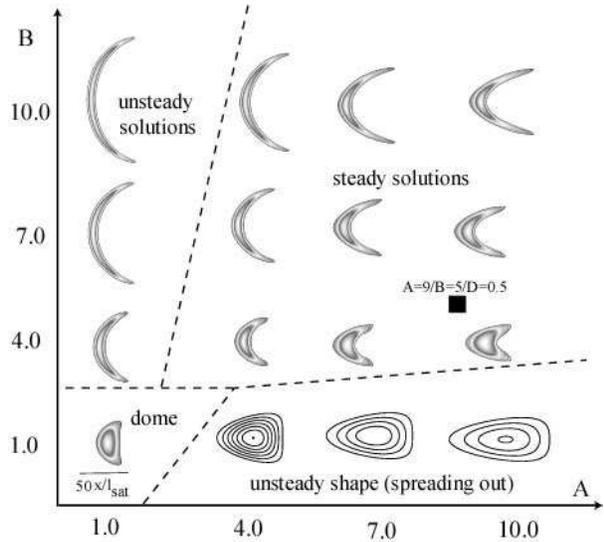} 
\caption{Evolution of the same initial \emph{cosine bump} sand pile
($h(x,y) = \cos(2\pi x/W_{0})\cos(2 \pi y/W_{0})$ with $W_{0} = 30
l_{sat}$, $H_{0} = 3 l_{sat}$) changing the value of the parameters
$A$ and $B$.  $D$ is kept constant and equal to $0.5$.  The dashed
lines separate qualitatively the different stability and shape
domains.  The dome domain is divided into two parts depending on the
steadiness of the dome solution.}
\label{fig9}
\vspace{-5mm}
\end{figure}

However, looking only at the $D$ influence is not accurate enough to
describe qualitatively the influence of diffusion.  In fact, the shape
is the result of three physical effects governed by the parameters
$A$, $B$ and $D$.  Fig.~\ref{fig9} presents a phase diagram with $D$
kept constant at a value of $0.5$.  Many different morphologies can be
observed, from a large and thin unsteady crescent to a flat unsteady
sand patch, and with steady barchans inbetween.  First, $A$ governs
the stability of the sand bed, and then a large $A$ tends to flatten
the sand-pile.  Increasing $B$, on the other hand, forces the slope to
increase and to nucleate a slip-face.  Thus, when the ratio $A/B$ is
high, there is hardly a slip face.  On the contrary, when it is low,
the slip face develops on the whole dune.  Thus the $A/B$ parameter
controls the way the slip-face appears.  Secondly, one can observe
that the dunes obtained with a constant ratio $A/B$ have different
morphologies.  This shows that for small $A$ and $B$ the coupling
coefficient $D$ plays an important role in the shaping of the dune. 
Field measurements of the horns' width and the minimal size of
barchans should help us to give an estimation of these parameters. 
The following table provides with a summary of the different
qualitative effects of $A$,$B$ and $D$.
\begin{center}
\begin{tabular}{lll}
\hline \hline
$~$      & physical mechanism & Main effect when increasing \\
\hline \hline
$A$      & curvature effect   &  $L$ increases,\\
         &                    &  $W$ decreases. \\
$B$      & slope effect       &  Slip-face appears easily, \\
         &                    &  $\dr_{x} h$ increases, $W$ increases.  \\
$D$      & Lateral coupling   &  $H$ decreases,  \\
         &                    &  $W$ and horns width increase,\\
	 &                    &  output flux increases.\\   
\hline \hline
\end{tabular}
\end{center}
Finally, the initial sand-pile does not always reach a constant shape. 
For large $A/B$ no steady state appears.  This is due to the fact that
the erosion is so strong that it cannot be balanced by the influx, and
the sand-pile just becomes flatter and flatter.  On the contrary, for
a large $B/A$, a thin unsteady crescent appears.  Finally, for a large
$D$ (compared to $A$ and $B$) steady and unsteady dome solutions are
produced by the numerical model.  This shows the importance of $A$,
$B$ and $D$ on the shape and on the minimal size of computed barchan
dunes.
%

\section{Conclusion}

We have seen that reptation can be used to describe the $3d$
crescentic shape of barchan dunes.  As {\it reptons} are sensitive on
the local slope, they induce a lateral sand flux.  This allows a given
sand-pile blown by the wind to reach an equilibrium shape, despite
varying "slices" speed and intrinsic differences between the main body
and the horns.  The crescentic shape comes from two mechanisms, which
compete with each other.  The first one is the speed dispersion within
the dune: the small height slices are faster than higher ones, leading
to the crescentic shape, and eventually to the destruction of the
barchan dune.  The second one, is the lateral flux deflection that
reduces the horns' speed by decreasing the erosion flux on them.  It
leads to the homogenization of the speed of the different slices, and
eventually to the propagation of a steady state dune or a steady sand
dome.  This mechanism is efficient since the overall excess flux can
escape from the horns.  In other words, barchanic shape appears to be
the very basic shape taken by any sand-pile blown by a unidirectional
wind.  Now that the steady shape of barchan dunes seems to be well
described, it would be interesting to study the dynamics of such a
structure placed far from equilibrium conditions, because, in the
field, input flux has no reason to be related to the output flux of a
dune.

As a matter of fact, in this paper, barchans have been numerically
obtained with quasi-periodic boundary conditions.  If standard
boundary conditions are used, a typical barchan dune takes the
crescentic shape, for the same reasons as the ones exposed in this
article, but it keeps growing or shrinking depending on the incident
sand flux, either higher or lower than the sand flux escaping from the
horns\cite{HACHD03}.  This point is crucial because it suggests that
single real barchan dunes are not steady structures and, moreover, it
shows that a crescentic shape can form even if the dune is not in an
equilibrium state.  Therefore, further investigations are still
required to understand the long-term existence and shape of barchan
dune in the field.  In our model we have neglected saltation coupling;
but we believe that it is not the dominant effect.  Moreover, it could
easily be taken into account by adding a new coupling coefficient
dedicated to lateral deflection of {\it saltons}.

Finally, it appears that depending on the coupling coefficient, $D$,
the solution can either be a $3d$ dome or a $3d$ barchan.  Similarly,
changing $A$ and $B$ can lead to different shapes, and different
behavior.  it might be interesting to understand the transformation of
a $3d$ sand patch, steady or unsteady, into a barchan dune because of
fluctuations of these parameters, and particularly of $D$ variations. 
As a matter of fact, these parameters may depend on external
conditions such as grains size distribution, humidity or wind
fluctuation.

{\bf Acknowledgement} : The author is grateful to T. Bohr for his
constructive remarks.  The author whishes to thank K.H. Andersen, B.
Andreotti, P. Claudin and S. Douady for their important role in this
work, and also S. Bohn for delightful discussions about barchans. 
This work has benefited from an ACI J.Ch.


\begin{thebibliography}{10}
\bibitem{Rque3} These boundary conditions induces also a weak coupling
between slices: the output flux is not homogeneous, so a part of the
sand escaping from the horns is transfered to the center of the dune.
\end{thebibliography}
\end{document}